\documentclass[epj]{svjour}
\usepackage[T1]{fontenc}
\usepackage[latin2]{inputenc}
\usepackage[english]{babel}
\usepackage[pdftex]{graphicx}
\usepackage{amssymb,amsmath}
\usepackage{wrapfig}
\newcommand{\parenth}[1]{\left( #1 \right)}

\newcommand{\abs}[1]{\left\lvert #1 \right\rvert}

\newcommand{\ket}[1]{\left\lvert \left. #1 \right\rangle \right.}

\newcommand{\scalprod}[2]{\left\langle #1 \vert #2 \right\rangle}

\begin{document}
\title{Role of the laser chirp parameter in photoionisation}
\author{Mih\'{a}ly Andr\'{a}s Pocsai\inst{1,}\inst{2} \and Imre Ferenc Barna\inst{1,}\inst{3}
}                     
%
%
\institute{Wigner Research Centre for Physics of the Hungarian Academy of Sciences 
\\ Konkoly--Thege Mikl\'os \'ut 29--33, H-1121 Budapest, Hungary
\\ \email{pocsai.mihaly@wigner.mta.hu} \and University of P\'ecs, Institute of Physics, Ifj\'us\'ag \'utja 6, H-7624 P\'ecs, Hungary \and ELI-HU Nonprofit Ltd.\\
Dugonics T\'er 13, H-6720 Szeged, Hungary}
\date{Received: date / Revised version: date}
%
\abstract{
Nowadays the development of novel particle accelerators is a hot topic for both experimental and theoretical sciences. In the CERN-AWAKE experiment electrons are accelerated in a cold rubidium plasma, generated by a short, intense laser pulse. We describe the corresponding photoionisation process with \textit{ab initio} quantum mechanical calculations. In this paper we primarily investigate how the frequency chirping of the applied laser pulse influences the photoionisation. Secondarily we also study how the results of the numerical simulations depend on the shape of the envelope of the laser pulse. Based on our results, we give a recommendation to fine-tune the physical parameters of the applied laser pulse in order to improve the homogeneity of the rubidium plasma.
} 
\maketitle
\section{Introduction}
\label{intro}
Laser and plasma based particle accelerators are becoming more and more important during the last decades in experimental physics. The original idea, proposed by Tajima and Dawson \cite{Tajima} promised a scheme for electron acceleration that provides the opportunity of constructing compact electron accelerators that are much cheaper to construct and operate compared to the conventional particle accelerators that implement the well-known storage ring technology. Tajima and Dawson recognized that short laser pulses with properly selected parameters create plasma waves that are eligible to accelerate short electron bunches. In this scheme, the magnitude of the accelerating field can be even thousand times larger than the maximum that is  achievable in conventional particle accelerators. Furthermore, as laser and plasma based accelerators are linear accelerators, the electron beams don't suffer severe energy loss due to synchrotron radiation. However, in the early 80's it was not possible to generate high intensity laser pulses. Therefore it was not possible to experimentally demonstrate the extremely high accelerating fields. This problem was solved by Strickland and Mourou \cite{cpa2} as they proposed a scheme, called chirped pulse amplification, to generate intense laser pulses without damaging the amplifier medium. Their idea is still the basis of modern laser systems. Esarey \textit{et.~al} excellently summarize the application of short laser pulses for plasma based electron acceleration in their review \cite{Esarey2}.

In the CERN-AWAKE experiment a short, high intensity infrared laser pulse is applied to generate plasma from rubidium, which serves as the accelerator medium for the planned electron accelerator. The detailed description of the proposal can be found in References \cite{awake,AWAKE2,Caldwell}. Our study connects to this experiment by describing the corresponding photoionisation phenomena with \textit{ab initio} quantum mechanical simulations. In our earlier paper we already investigated some basic properties, like the probability of photoionisation or the photoelectron energy spectrum at different laser intensities \cite{pocsai-rb}. Continuing our earlier work, now we investigate primarily the variation of the energy spectrum corresponding to application of frequency chirp to the ionising laser pulse. Our simulations are based on the time-dependent close coupling method, originally developed by Bray \cite{bray}. Chatel \textit{et.~al} discovered that in excited sodium vapor, direct and sequential paths of two-photon ionisation compete with each other at certain values of frequency chirp, creating interesting interference patterns. Balling \textit{et.~al} showed that for some special values of chirp the probability of some atomic transitions of rubidium atoms are highly enhanced \cite{Balling}. They also proposed an experimental method to implement frequency chirp. This experimental realisation may be important for the CERN-AWAKE experiment: in this paper we show that properly chirped ionising laser pulses may improve the homogeneity of the rubidium plasma, which is a crucial point in that experiment.

\section{Overview of the applied theory}
\label{sec:theory}

To describe photoionisation, we apply \textit{ab initio} quantum mechanical calculations. In our model we apply the one active electron approximation, i.e.~the contribution of the inner shell electrons is taken into consideration via shielding the valence electron from the atomic core. The time-dependent Schr\"{o}dinger-equation refers to the wave function of the valence electron:

\begin{equation}\label{eq:TDSE}
	i \partial_{t} \ket{\Psi (t, \mathbf{r})} = \hat{H} \ket{\Psi (t, \mathbf{r})}.
\end{equation}
with
\begin{equation}
	\hat{H} = \hat{H}_{Rb} + \hat{V}_{I}
\end{equation}
In the Hamiltonian operator above,
$\hat{H}_{Rb}$ is the Hamiltonian operator of the free rubidium atom and $\hat{V}_{I}$ describes the interaction of the rubidium atom with the external laser field. Atomic units are used throughout the paper unless stated otherwise.

For one active electron approximation, there are many well-known model potentials that have been applied successfully, see e.g.~\cite{green,garvey,garvey2}. To describe the free rubidium atom, we chose the Hellmann pseudopotential, described in Ref.~\cite{milosevic}:
\begin{equation}\label{eq:Hamiltonian}
	\hat{H}_{Rb} = -\frac{1}{2} \nabla^{2} - \frac{1}{r}(1 - b e^{-dr}).
\end{equation}
For rubidium, also listed in Ref.~\cite{milosevic}, the shielding parameters $b$ and $d$ take the values $b = 4.5$ and $d = 1.09993$. This model potential has the big advantage against Green's one that many matrix elements can be calculated analytically due to its simple form, requiring significantly less computational efforts in the end.

Expanding the solution of the TDSE on the basis of the eigenfunctions of the free Hamiltonian operator, applying time-dependent expansion coefficients, it reduces to a system of linear ordinary differential equations, also called coupled channel equations:
\begin{equation}\label{eq:coupledtransformed}
	i \dot{a}_{k}(t) = \sum_{j=1}^{N} V_{kj} a_{j}(t) + E_{k} a_{k}(t).
\end{equation}
with $a_{k}, \, k = 1 \dots N$ being the time-dependent expansion coefficients mentioned above, each of them corresponding to a possible state of the rubidium atom, i.e.~an eigenfunction of $\hat{H}_{Rb}$. A typical situation is that the rubidium atom is in its ground state before the interaction, i.e.
\begin{equation}
	\begin{aligned}
a_{k} \left( t \rightarrow - \infty \right) =
\end{aligned}
\left\{
\begin{aligned}
& 1  \quad & k=1   \\
& 0 \quad & k \neq 1
\end{aligned} 
\right.
\end{equation}
Of course, an arbitrary mixture of the states can be chosen as initial condition.

The final state occupation probability for each state reads:
\begin{equation}
	P_{k} (t \rightarrow \infty) = \abs{a_{k} (t \rightarrow \infty)}^{2}
\end{equation}
The (photoelectron) can also be easily calculated:
\begin{equation}
	\frac{\partial P}{\partial E} = \sum_{l} \abs{\scalprod{\Phi_{E}^{l}(\mathbf{r})}{\Psi(t=T,\mathbf{r})}}^{2}
\end{equation}
where $\Phi_{E}^{l}(\mathbf{r})$ is the corresponding continuum state with energy $E$ and azimuthal quantum number $l$ and $\Psi(t \rightarrow \infty, \mathbf{r})$ is the final state wave function.

The bound states of the valence electron have been described with Slater-type orbitals
\begin{equation}\label{eq:slater}
	\chi_{n,l,m,{\kappa}}(\vec{r}) 
= C(n,{\kappa})r^{n-1}e^{-{\kappa}r}Y_{l,m}    ({\theta},{\varphi})
\end{equation}
Here, $n,l$ and $m$ are the principal, azimuthal and magnetic quantum numbers, respectively. $\kappa$ is the screening constant. Its value depends on the basis that has been chosen for the calculations, and it also determines the calculated energy eigenvalue of the corresponding bound state. The normalisation factor of the Slater orbitals reads:
\begin{equation}
	C \parenth{n, {\kappa}} = \frac{\parenth{2 \kappa}^{n+1/2}}{\sqrt{\parenth{2n}!}}.
\end{equation}
Here we note that Slater orbitals with different principal quantum numbers are not orthogonal to each other.

For the description of the continuum states, we chose Coulomb wave packets:
\begin{equation}\label{eq:coulomb}
	\varphi_{k,l,m,\tilde{Z}}(\vec{r}) = N(k,\Delta k) 
 \int\limits_{k-\Delta k/2}^{k+\Delta k/2} 
  F_{l,\tilde{Z}}(k',r)dk'Y_{l,m}({\theta},{\varphi})
\end{equation}
They are the probability amplitudes of an electron with energy $\varepsilon$ such that $E - \Delta E < \varepsilon < E + \Delta E$. In the equation above $k$ and $\Delta k$ are the momentum and momentum width corresponding to the continuum energy and energy width $E$ and $\Delta E$, $F_{l,\tilde{Z}}(k,r)$ is the Coulomb wave function with $l$ and $m$ being the azimuthal and magnetic quantum numbers, respectively and $\tilde{Z}$ the charge of the ion. $\tilde{Z} = 1$ corresponds to single ionisation. The Coulomb wave packets with non-overlapping energy ranges are orthogonal to each other.Their normalization factor reads:
\begin{equation}
	N(k, \Delta k) = \frac{1}{\sqrt{k \Delta k}}.
\end{equation}
The Coulomb wave function has the form of
\begin{equation}\label{eq:coulomb_wf}
\begin{split}
	F_{l,\tilde{Z}} (k,r) = &\sqrt{\frac{2k}{\pi}} \exp \parenth{\frac{\pi \tilde{Z}}{2k}} \frac{(2kr)^{l}}{(2l+1)!} \exp \parenth{-ikr} \times\\
	& \abs{\Gamma(l+1-i \tilde{Z}/k)} \times\\
	& _{1}F_{1}(1+l+i \tilde{Z}/k, 2l+2, 2ikr).
\end{split}
\end{equation}

To describe the interaction of the rubidium atom with the external laser field, we write interaction potential in length gauge:
\begin{equation}
	\hat{V}_{I} = \mathbf{r} \cdot \mathbf{E}(t, \mathbf{r}).
\end{equation}
We also took into consideration that in our scope, the dipole approximation is valid. In general, the electric field has the form of
\begin{equation}
	\mathbf{E}(t) = \boldsymbol{\varepsilon} E_{0} f(t) \sin \parenth{\omega_{L}(t) t}
\end{equation}
Here $\boldsymbol{\varepsilon}$ is the polarization vector, $E_{0}$ the amplitude, $f(t)$ the envelope and $\omega_{L} (t)$ the frequency the laser pulse. Note that the frequency may explicitly depend on the time. Laser pulses with a given time dependence for their frequency are called frequency chirped laser pulses. Due to practical purposes, we chose linear frequency chirp, i.e.
\begin{equation}
	\omega_{L} (t) = \omega_{0} + \sigma t
\end{equation}
where, depending on the shape of $f(t)$, $\omega_{0}$ is the initial (e.g.~for $f(t)$ being the sine square function) or central (e.g.~for $f(t)$ being a cosine square or a Gaussian function) frequency and $\sigma$ being the chirp parameter. For positive or negative values of the chirp parameter, one says by definition that, the laser pulse is positively or negatively chirped, respectively.

For a more detailed theoretical description, see Ref.~\cite{bethe-saltpeter} and our earlier work, Ref.~\cite{pocsai-rb}.

\section{Results}
\label{sec:results}
In our earlier work we have already shown that the method described in the previous section can be successfully applied to investigate photoionisation phenomena for rubidium atoms. We reproduced the measured bound state energy spectrum within a few per cent accuracy, showed that at relevant laser intensities the probability of photoionisation is nearly $100 \, \%$, reproducing the saturation phenomenon and also calculated the photoelectron energy spectrum for different intensities \cite{pocsai-rb}. The ATI peaks were clearly visible on those spectra. Here we continue our work and investigate how the positive or negative chirp influences the energy spectrum. Furthermore we also investigate how the shape of the envelope influences the results of the numerical simulation.

\subsection{Application of frequency chirped laser pulse}
Applying frequency chirp to the laser pulse modifies the photoelectron energy spectrum. We set the chirp parameter such that the frequency of the laser pulse varies $10 \, \%$ of the initial laser frequency pro pulse duration. The other laser parameters were $\lambda = 800 \, \mathrm{nm}$, $T = 120 \, \mathrm{fs}$ and $I = 8 \cdot 10^{10} \, \mathrm{cm}^{-2}$. We chose the laser intensity relatively small for demonstration purposes, that is, we chose the laser intensity such that the AIT peaks are not as ``noisy'' as the ones presented in Ref.~\cite{pocsai-rb}.

\begin{figure}
\begin{center}
	\includegraphics[width=0.45\textwidth]{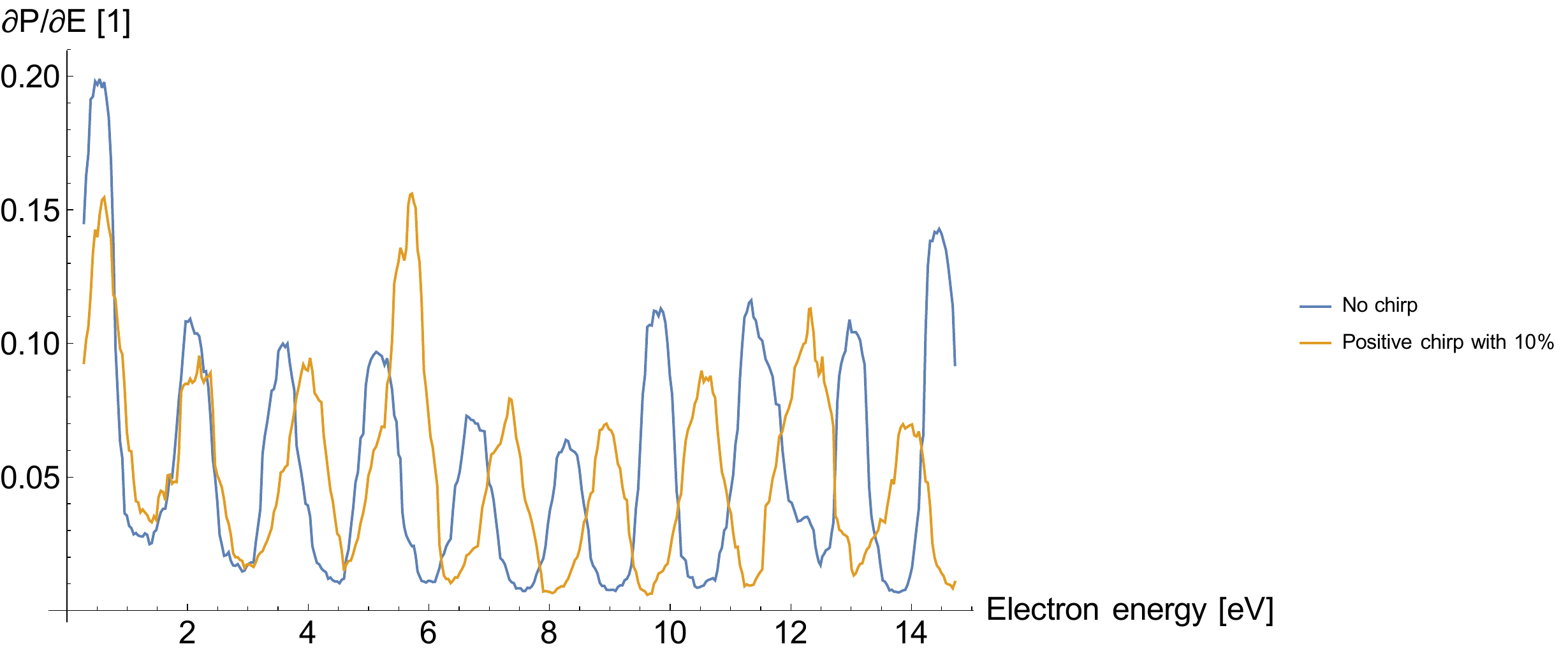}
	\caption{Photoelectron energy spectrum with (orange line) and without (blue line) applying frequency chirp to the laser pulse. As the chirp parameter is positive, the distance between the ATI peaks increases, meaning that the average kinetic energy of the photoelectrons increases accordingly.}
	\label{fig:spect_pos_chirp}
\end{center}
\end{figure}

\begin{figure}
\begin{center}
	\includegraphics[width=0.45\textwidth]{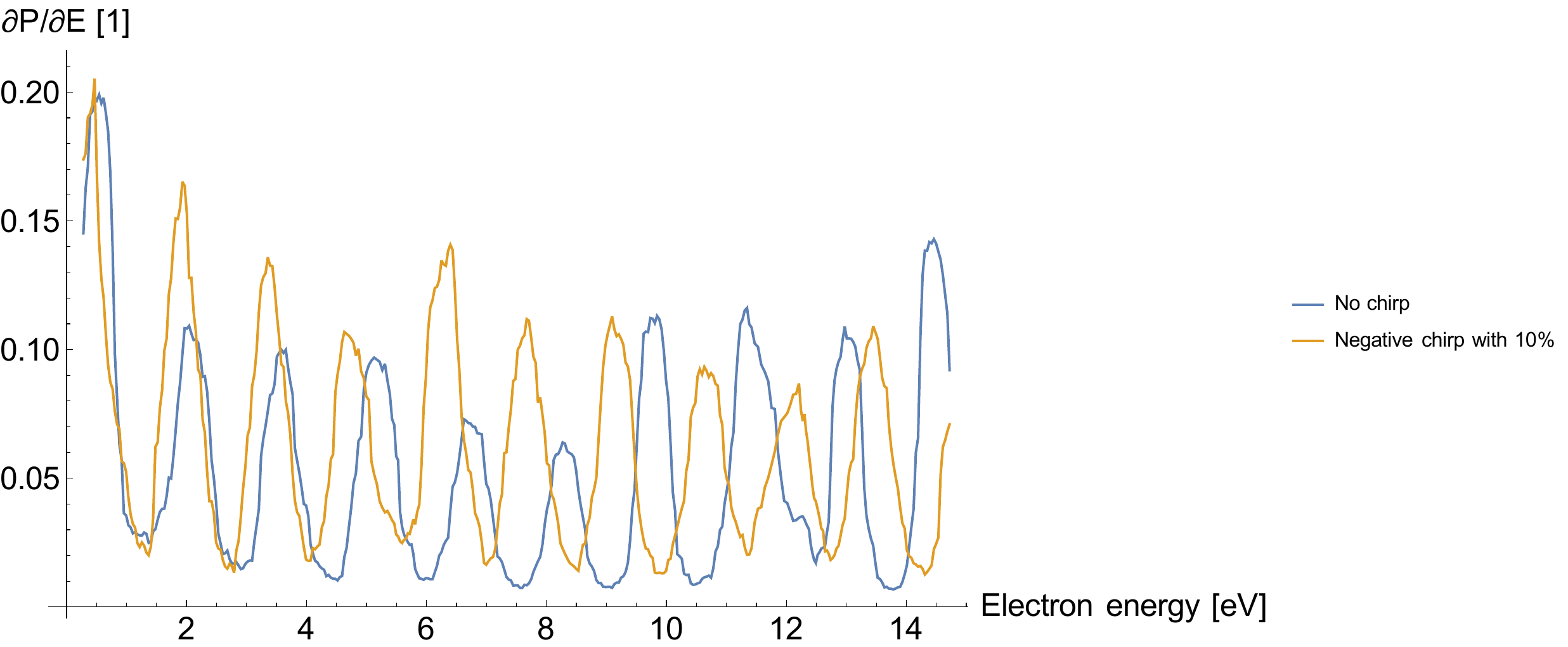}
	\caption{Photoelectron energy spectrum with (orange line) and without (blue line) applying frequency chirp to the laser pulse. As the chirp parameter is negative, the distance between the ATI peaks increases, meaning that the average kinetic energy of the photoelectrons decreases accordingly.}
	\label{fig:spect_neg_chirp}
\end{center}
\end{figure}

Figures \ref{fig:spect_pos_chirp} and \ref{fig:spect_neg_chirp} show that the distance between the ATI peaks, and therefore the average photoelectron kinetic energy slightly increases for positive values of the chirp parameter and decreases for negative values, compared to the photoelectron energy spectrum corresponding to a laser pulse without frequency chirp. This suggests that applying negatively chirped laser pulses for generating the rubidium plasma will result in a plasma with less collisions within it, comparing to the one generated by a non-chirped laser pulse. That is, less average photoelectron kinetic energy and therefore less collisions in the plasma will result in a more homogeneous plasma density.

\subsection{The envelope of the laser pulse}
To investigate the influence of the envelope on the numerical results, we compared the photoionisation probability and photoelectron energy spectra calculated with cosine square and Gaussian envelope as the envelope
\begin{equation}
	f(t) = \cos^{2} \parenth{\frac{t}{T}}
\end{equation}
is a good approximation of the envelope
\begin{equation}
	g(t) = \exp \parenth{-\frac{t^{2}}{T^{2}}}.
\end{equation}
In this sense, good approximation refers to the fact that expanding both $f(t)$ and $g(t)$ into a Taylor-series, they are equal up to the second order term. The cosine square function has two advantages, compared with the Gaussian: first, it has a compact support, i.e.~it is exactly zero for $\abs{t/T} \geq \pi / 2$, whilst the Gaussian envelope tends to zero only as $t \rightarrow \pm \infty$, therefore it minimizes the numerical errors during calculation. Second, one has to integrate the Schr\"{o}dinger equation---or equivalently, the coupled channel equations---on a shorter time interval that results in (significantly) shorter computation time in the end. On the other hand, the cosine square envelope is ``artificial'' and Gaussian is ``natural'' in the sense that Gaussian waves are approximate solutions of Maxwell's equations within the confines of paraxial approximation and therefore have a high practical importance.

\begin{figure}
\begin{center}
	\includegraphics[width=0.45\textwidth]{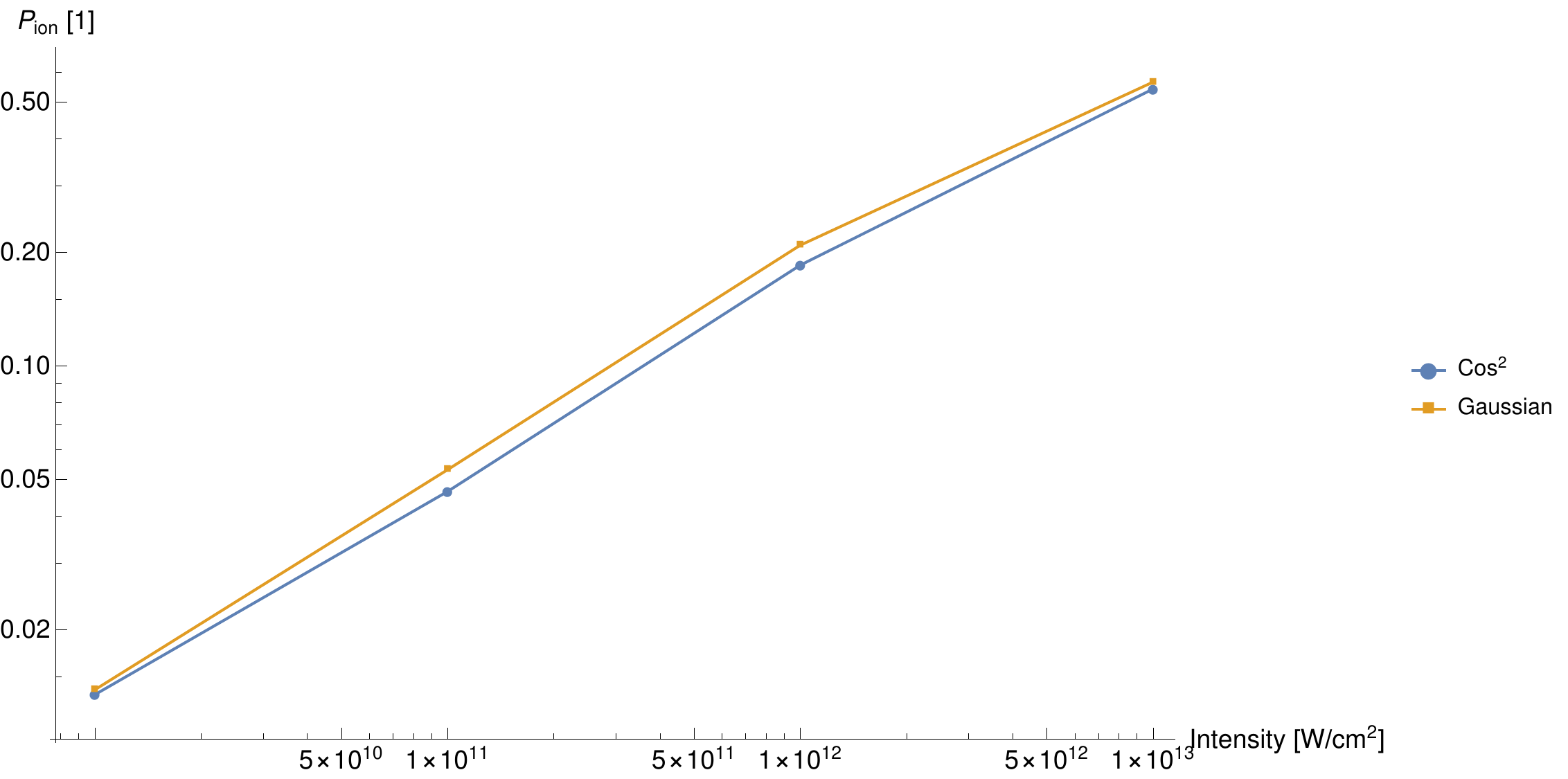}
	\caption{Comparison the photoionisation probability as a function of laser intensity for cosine square shaped (blue curve) and Gaussian (orange curve) envelopes. The laser intensity varies between $10^{10} \mathrm{W} \cdot \mathrm{cm}^{-2}$ and $10^{13} \mathrm{W} \cdot \mathrm{cm}^{-2}$.}
	\label{fig:compare-probs}
\end{center}
\end{figure}

\begin{figure}
\begin{center}
	\includegraphics[width=0.45\textwidth]{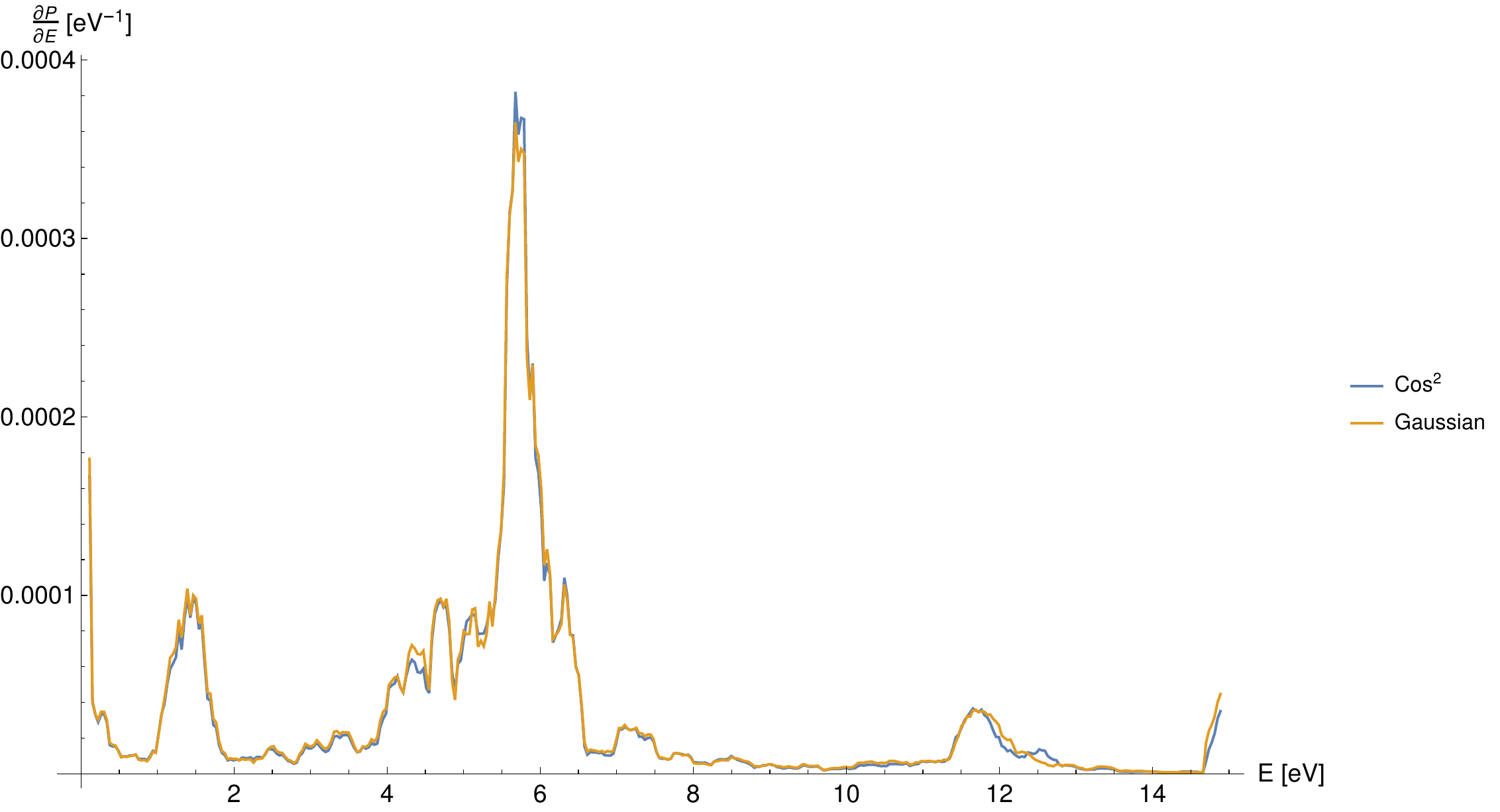}
	\caption{Comparing the photoionisation probability as a function of laser intensity for cosine square shaped (blue curve) and Gaussian (orange curve) envelopes with laser intensity $I = 10^{10} \mathrm{W} \cdot \mathrm{cm}^{-2}$}
	\label{fig:compare-spectra-1}
\end{center}
\end{figure}

\begin{figure}
\begin{center}
	\includegraphics[width=0.45\textwidth]{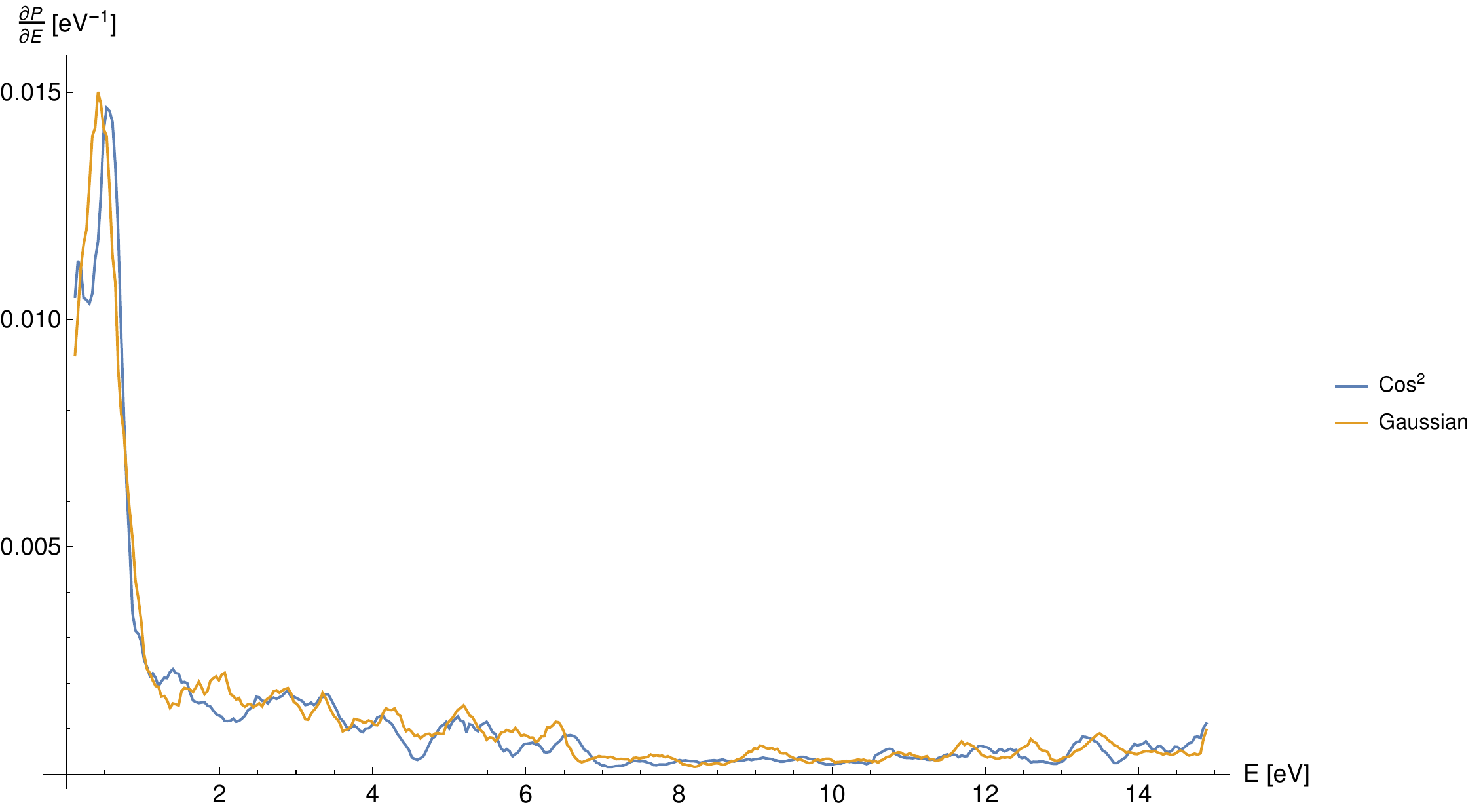}
	\caption{Comparing the photoionisation probability as a function of laser intensity for cosine square shaped (blue curve) and Gaussian (orange curve) envelopes with laser intensity $I = 10^{13} \mathrm{W} \cdot \mathrm{cm}^{-2}$}
	\label{fig:compare-spectra-2}
\end{center}
\end{figure}

For demonstration purposes, we chose the pulse duration to be short, $T = 2.5 \, \mathrm{fs}$. Again, the -wavelength has been set to $\lambda = 800 \, \mathrm{nm}$. We found that the cosine square shaped envelope slightly underestimates the photoionisation probability compared to the Gaussian envelope, see fig.~\ref{fig:compare-probs} This difference can be easlily understood as the Gaussian envelope tends to zero only as $t \rightarrow \pm \infty$, whilst the cosine square envelope has a compact support, i.e.~it is exactly zero for $\abs{t/T} \geq \pi / 2$. In the photoelectron energy spectrum, there is almost no difference between the two cases for low intensities. This small difference gets larger as the intensity increases, see figures \ref{fig:compare-spectra-1} and \ref{fig:compare-spectra-2}. However, even at higher intensities, the difference is still negligible. This suggests that the cosine square shaped envelope is eligible for predictions corresponding to photoionisation.

\section{Conclusion}
In the present paper we investigated how the frequency chirp of the laser pulse influences the photoelectron energy spectrum and also, how the shape of the envelope of the laser pulse influences predictions obtained from the numerical simulations.

For the first case, we found that, positively chirped laser pulses increase and negatively chirped laser pulses decrease both the distance of the ATI peaks, which is approximately $1.55 \, \mathrm{eV}$, i.e.~one photon energy for the non-chirped case, and also the average photoelectron kinetic energy. Lower average photoelectron kinetic energy suggests more homogeneous plasma density. Therefore we recommend to apply negatively chirped laser pulses in the CERN-AWAKE experiment and fine-tune the chirp parameter according to the experimental results.

For the second case, we found for the envelope of the laser pulse that even though a cosine square shaped envelope is artificial in some sense, it is still eligible for realistic predictions as it is a good approximation of the Gaussian function. Furthermore, it has the advantage that the required computational capacity may be significantly less than that of corresponding to the Gaussian envelope, and it also produces also less numerical errors due to its compact support.

\section*{Acknowledgement}
\begin{wrapfigure}[3]{l}{0.1\textwidth}
	\vspace*{-22pt}
	\includegraphics[width=0.095\textwidth]{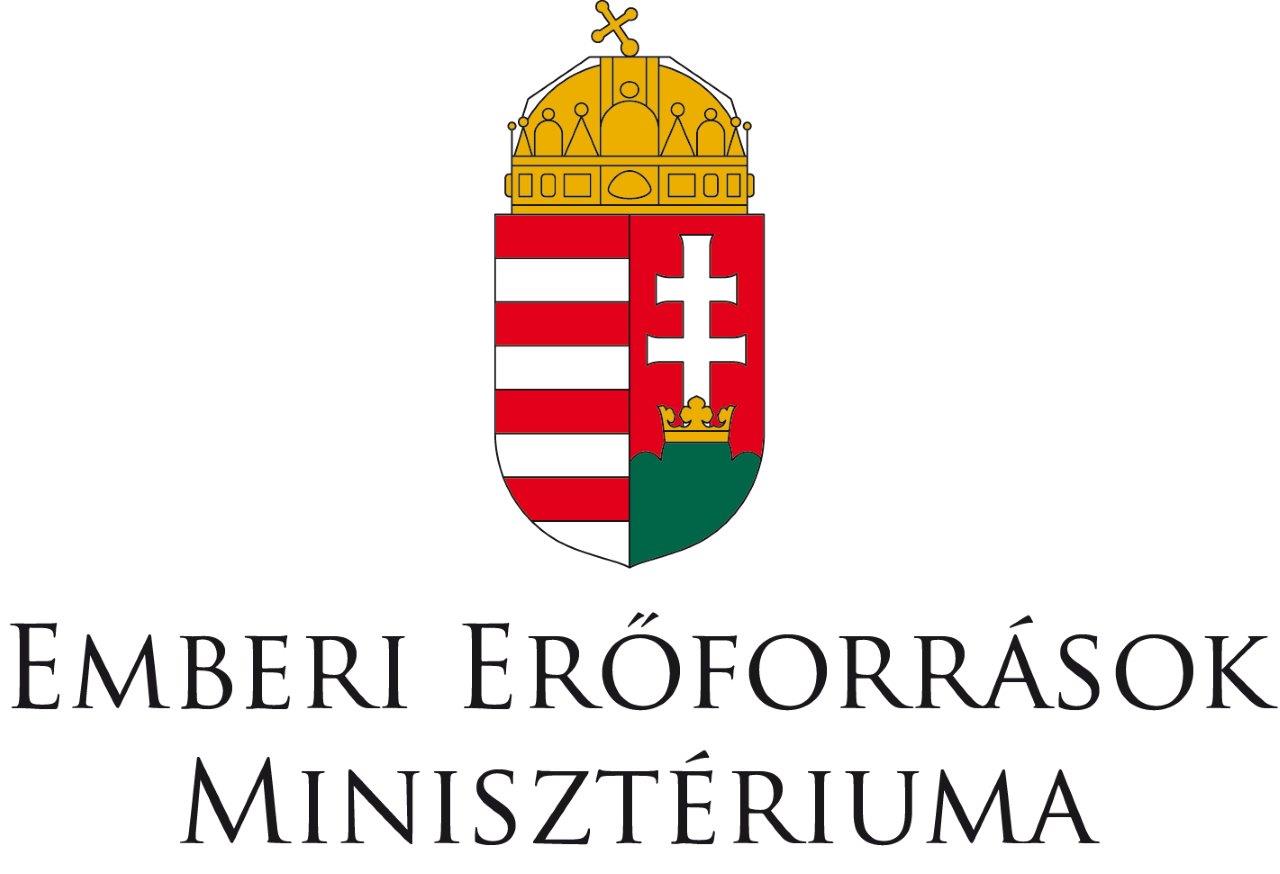}
\end{wrapfigure}

This study was supported By the \'{U}NKP-18-3 New National Excellence Program of the Ministry of Human Capacities. Author M.~A.~Pocsai would like to acknowledge the support by the Wigner GPU Laboratory of the Wigner RCP of the H.A.S.

\bibliographystyle{epj}
\bibliography{citations}

\end{document}